\documentclass{article}

\usepackage{PRIMEarxiv}

\usepackage[utf8]{inputenc} 
\usepackage[T1]{fontenc}    
\usepackage{hyperref}       
\usepackage{url}            
\usepackage{booktabs}       
\usepackage{amsfonts}       
\usepackage{nicefrac}       
\usepackage{microtype}      
\usepackage{lipsum}
\usepackage{fancyhdr}       
\usepackage{graphicx}       
\graphicspath{{media/}}     

\title{Translating Place-Related Questions to GeoSPARQL Queries}
\author{
  Ehsan Hamzei, Martin Tomko, Stephan Winter \\
  The University of Melbourne \\
  Parkville\\
  \texttt{\{ehsan.hamzei, tomkom, winter\}@unimelb.edu.au} \\
}

\usepackage{xspace}
\newcommand{\ex}[1]{\textit{#1}\xspace}

\newcommand{\pattern}[1]{\texttt{#1}\xspace}
\usepackage[square,numbers]{natbib}
\usepackage{makecell} 

\usepackage{listings}
\usepackage{algorithm}

\newdimen\longformulasindent
\newenvironment{longformulas}
 {\global\longformulasindent=0pt
  \def\>{\global\advance\longformulasindent2em\relax\hspace{2em}}%
  \def\<{\global\advance\longformulasindent-2em\relax\hspace{-2em}}%
  \begin{array}{@{}>{\displaystyle\hspace{\longformulasindent}}l@{}}}
 {\end{array}}

\begin{document}

\maketitle



\begin{abstract}
Many place-related questions can only be answered by complex spatial reasoning, a task poorly supported by factoid question retrieval. Such reasoning using combinations of spatial and non-spatial criteria pertinent to place-related questions is increasingly possible on linked data knowledge bases. Yet, to enable question answering based on linked knowledge bases, natural language questions must first be re-formulated as formal queries. Here, we first present an enhanced version of YAGO2geo, the geospatially-enabled variant of the YAGO2 knowledge base, by linking and adding more than one million places from OpenStreetMap data to YAGO2. We then propose a novel approach to translate the place-related questions into logical representations, theoretically grounded in the \emph{core concepts of spatial information}. Next, we use a dynamic template-based approach to generate fully executable GeoSPARQL queries from the logical representations. We test our approach using the Geospatial Gold Standard dataset and report substantial improvements over existing methods.
\end{abstract}


\section{Introduction}
People frequently search on the Web seeking answers to pragmatic information needs, such as evaluating the number of facilities (i.e., pharmacies) in a suburb where they may want to move to. Such place-related questions often require the use of structured Web content (linked data).
Consider the following question, taken from the Geospatial Gold Standard dataset \cite{Punjani:2018}:

\vspace{0.1cm}
\textbf{Question}: \textit{How many pharmacies are in 200 meter radius of High Street in Oxford?}
\vspace{0.1cm}

\noindent Structured data stored in databases and knowledge bases are more suitable than unstructured text retrieval to answer such questions. Using structured data, Geographic Question Answering (GeoQA) systems can perform spatial (i.e., 200 meter radius) and non-spatial operations (i.e., count \emph{how many}), and retrieve information based on specific criteria (e.g., \emph{High Street in Oxford}).

The complementary role of structured data for Question Answering (QA) systems has been emphatically noted \cite[e.g.,][]{dimitrakis2019survey, Chen:2014}, yet answering natural language questions using structured geospatial data remains challenging. Similar to open-domain QA systems, GeoQA systems require the ability to generate structured queries from natural language questions. However, vagueness in place types both in terms of their boundaries (e.g., \emph{downtown}), meaning (e.g., \emph{bakeries} vs. \emph{cafes}), and relations (e.g., \emph{near}) makes this task significantly more difficult \cite{RN43,RN26}.

Translating natural language questions to formal queries often involves two major steps: (1) parsing the questions into an intermediate structure, and (2) generating queries using the intermediate structure \cite{Punjani:2018}. In the first step, concepts such as place names are extracted from the natural language questions, and their relations are expressed in graph or tree data structures. In the second step, the extracted concepts are used to define variables, and their relations are translated into the targeted structured language(s) based on their predefined syntax.

A parsing method in a domain-specific QA should be linked to the domain concepts to ease the query generation step. In this paper, the object-based conceptualization of place \cite{purves:2019,kuhn2021semantics} is used as the grounding to design a parsing method for place-related questions. We extend previous studies on analyzing place-related questions \cite{agile:2019, agile-giss-1-23-2020} to capture the extracted concepts and relations. We use the state-of-the-art language models to identify concepts and their relations. We capture the results of the parsing step in a logical representation that is both machine- and human-readable. Next, we devise a dynamic approach to translate the parsed questions to GeoSPARQL queries. The dynamic approach uses templates for partial GeoSPARQL \textit{constructs}, e.g., defining a concept, rather than a template for the \textit{whole query}. We show how our approach leads to significant improvements in translating questions to GeoSPARQL queries in comparison to the previous works. 

The novelty of the proposed method is twofold: (1) our method is grounded in geographic domain knowledge. Thus, instead of trying to fit a model on a dataset, we use our conceptualization to determine how concepts relate; (2) our method is reusable to translate the questions to other structured languages(s) with minimal efforts using logical representation that formally captures the gist of the place-related questions. In short, the paper:
\begin{itemize}
    \item Improves existing methods in translating natural language question to GeoSPARQL queries over the Geospatial Gold Standard dataset \cite{Punjani:2018}.
    \item Proposes an intermediate logical representation using the available domain knowledge that can be used to translate question to structured queries.
    \item Enriches an available knowledge base (YAGO2geo \cite{Punjani:2018}) by linking more than a million places from 500 place types using OpenStreetMap data.
    \item Uses the state-of-the-art language model (i.e., BERT embedding) to perform ontology mapping and evaluates their performance in mapping place types and properties.
\end{itemize}

\section{Related Works}
GeoQA is a sub-domain of Question Answering (QA) that focuses on generating answers to geographic questions \cite{ferres2017knowledge, agile-giss-2-8-2021}. Diverse information sources such as textual information \cite{Ferres:2006, mishra2010}, geodatabases \cite{chen2014}, and spatially-enabled knowledge bases \cite{Ferres:2010} have been investigated to enable GeoQA. Answers to geographic questions can then be presented in natural language, structured into tables and graphs, or visualised as maps \cite{Scheider:2020, Chen:2014, ferres2017knowledge}. 

Early research on the translation of geographic questions to structured queries links back to \cite{Zelle:1996} and \cite{Tang:2001}, who designed a method to capture geographic questions in logical form. This logical form captured factual statements extracted from the questions in terms of objects and predicates. This early work had limited generality, restricting the diversity of geographic questions that could be captured to only those relating to a narrow set of types of geographic places (e.g., \textit{cities}, \textit{countries}, \textit{states} and \textit{rivers}). The resulting formalism only offered a simplistic set of twenty properties and predefined relations (predicates), including the ability to define objects, get their properties such as population and area, and for defining spatial and logical relations such as \emph{capital-of} and \emph{equal} \cite{Zelle:1996, Tang:2001}.

Later, \cite{Chen:2014} devised a method to translate geographic questions into \textit{spatial SQL} queries. This method was only developed for four types of questions: (1) identifying coordinates of a place, (2) finding the distance between places, (3) finding the closest place to another place and (4) finding places in a predefined neighbourhood of another place. The method was based on annotated questions and predefined spatial SQL templates.

Recently, \cite{Punjani:2018} introduced a Gold Standard dataset for translating geographic questions to GeoSPARQL queries. They developed a template-based approach to generate GeoSPARQL queries from geographic questions. Their approach includes two steps of natural language processing (information extraction and entity resolution), and a query generator to parameterize GeoSPARQL templates. The initial work of \cite{Punjani:2018} was further refined by using a more comprehensive list of templates for geographic queries and diverse set of natural language processing toolkits \cite{punjani2021templatebased}. While small, the hand-curated set of questions proposed by \cite{Punjani:2018} remains the only evaluation Gold Standard dataset for GeoQA.

\cite{li2021ur} designed a parsing method using deep neural networks (DNN) to combine the preprocessing, information extraction, and relation identification steps into a DNN model. They produced GeoSPARQL queries out of the parsing results using a dynamic approach for query generation, yet the queries were not directly executable, as they lacked ontology mapping and concept identification. While recent studies in translating open-domain questions to structured queries \cite[e.g.,][]{xu2019nadaq,lyu2020hybrid,9207428} show that DNN models perform much better in comparison to traditional rule-based methods, the small Gold Standard dataset seems to be insufficient for training a DNN model. Consequently, the method proposed by \cite{punjani2021templatebased} performed better over the same dataset.

While several studies explored how to model geographic questions, a proper intermediate representation of the questions that utilizes the available domain knowledge is still missing. It is specially important for GeoQA because the available datasets are small and without a theoretical grounding the coverage of proposed methods cannot be evaluated. Here, we utilize available domain knowledge to propose an intermediate representation that captures semantics of questions without considering the technical features of a destination query language. We then use this intermediate representation to translate questions to GeoSPARQL queries. 

\section{Preliminaries}
Core concepts of spatial information, proposed by \cite{kuhn2012core}, include the sole base concept of \emph{location}, and a set of content and quality concepts. The content concepts are \emph{object}, \emph{field} and \emph{event}. A spatial object has an identity and is bounded in space (e.g., \emph{Mount Everest}), while a spatial field represents a geographic phenomenon that encompasses the whole space but its magnitude may differ from one location to another (e.g., \emph{terrain height}). Events are bounded not only in space but in time, and they may cause changes in spatial objects and fields (e.g., a \emph{hurricane}). Finally, the quality concepts determine the granularity and value of spatial information.

Using the core concepts of spatial information, geographic `places' are conceptualized as \textit{spatial objects with socially-constructed identities} \cite{purves:2019}. While geographic places include diverse types with varied and possibly heterogeneous characteristics \cite{doi:10.1080/13875868.2019.1688332}, this conceptualization captures geographic places at a high level of abstraction, without any bias or favor to specific place types. \cite{purves:2019} conceptualize places as spatial objects that:
\begin{itemize}
    \item may have associated \textit{properties}, and \textit{relations}, incl. spatial;
    \item have a \textit{location} and are bounded, yet their boundary may be fuzzy (e.g., \emph{downtown});
    \item may have `parts' or `aggregates';
    \item can participate in \emph{events}, and their properties or relation may be subjected to changes;
    \item can be carved out of fields (e.g., \emph{climate zones}).
\end{itemize}
Thus, to study place-related questions based on the object-based conceptualization, we must consider `places', their `location', other `properties', `relations', and also `events'.

\cite{agile:2019} proposed an encoding schema to analyze the content of place-related questions, which was later extended by \cite{agile-giss-1-23-2020} to increase its capabilities for analyzing a wider range of geographic questions. This schema captures the syntactic structure of the questions by labelling the tokens and phrases in natural language questions using a predefined set of encoding classes. The essential elements of this encoding schema are:

\begin{itemize}
    \item \textbf{place name} is a direct reference to a geographic place (e.g., \emph{New York}, \emph{Big Apple}).
    \item \textbf{place type} is a generic reference to a category of a taxonomy that captures places with similar functional, spatial and physical properties (e.g., \emph{mountain}).
    \item \textbf{properties} describe diverse characteristics of places, and a place may be described by a set of criteria imposed on these properties (e.g., \emph{population}, or \emph{area}).
    \item \textbf{activities} are afforded by places, and places may be queried for their affordances (e.g., [a place] \emph{to buy hardware}).
    \item \textbf{situations} are another way to describe places by reference to what is available or can be experienced there (e.g., [a place] \emph{to see birds}), instead of what one can do there.
    \item \textbf{qualities} can be a quality of an activity, a situation or a property of place that narrows down the search domain for identifying relevant places (e.g., the \emph{most populated city}, the \emph{old building}, or the \emph{best cafe}).
                \item \textbf{spatial relations} describe how places are located in a relative space, and includes a diverse set of topological (e.g., \emph{inside}), directional (e.g., \emph{north of}) and metric (e.g., \emph{in 200 meter}) relations \cite{ligozat13:qualitative}.
\end{itemize}

\section{Questions to Queries} 
\subsection{Dataset}
We use the Geospatial Gold Standard dataset \cite{Punjani:2018} to test the proposed translation method. This dataset contains 200 place-related questions collected for translating questions into GeoSPARQL queries. This is a handcrafted dataset generated by students of an Artificial Intelligence course. The participants were instructed to formulate questions about geographic places that can be answered using information available in knowledge bases or through spatial analysis supported in GeoSPARQL. The questions are about geographic places in the United Kingdom and The Republic of Ireland.

\cite{Punjani:2018} also provide a knowledge base that contains both thematic and spatial information about places in the UK and Ireland. They link detailed spatial information extracted from the OpenStreetMap (only natural features) and GADM (administrative levels) datasets to the YAGO knowledge base \cite{HOFFART201328} which contains accurate thematic information extracted from DBPedia, GeoNames and WordNet \cite{10.1145/2487788.2487935}.

We extended this dataset by adding more than one million places in the UK and Ireland that belong to roughly 500 place types to enrich the YAGO2geo that originally contains spatial information for natural and administrative places. We collected data using the OSM Overpass Turbo API\footnote{\url{http://overpass-turbo.eu/}}, and used the strategy of \cite{Punjani:2018} to link these spatial data to the YAGO knowledge base. The enriched dataset now contains accurate spatial information about point of interests (e.g., \emph{tourist attractions}), amenities (e.g., \emph{restaurants}), historic places (e.g., \emph{historic monuments}), buildings (e.g., \emph{schools}) and shops (e.g., \emph{boutiques})\footnote{OSM data keys: \url{https://wiki.openstreetmap.org/wiki/Category:Key_descriptions}}. Thematic information about places also extended by addresses, contact numbers and websites, wherever they are provided\footnote{Available at \url{https://github.com/hamzeiehsan/Questions-To-GeoSPARQL}}.

\subsection{Method} 
\begin{figure}[]
\includegraphics[width=\linewidth]{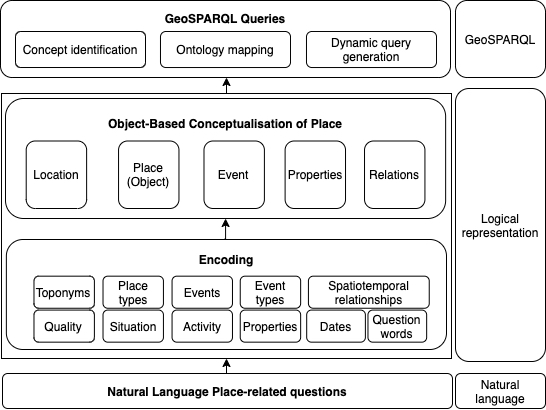}
\caption{Method workflow to translate place-related question to GeoSPARQL queries.}
\label{fig:chap5method}
\end{figure}

Figure~\ref{fig:chap5method} shows the workflow of the proposed method to translate place-related questions to structured queries\footnote{Demo is available at: \url{https://tomko.org/demo/}}. The workflow starts with extracting encodings to identify and annotate place-related semantics. Next, the relations among the encodings are identified through grammatical parsing. Then, the encodings and their relationships are expressed in logical statements. Finally, the logical statements are translated into GeoSPARQL queries.


\subsection{Encoding Extraction}
The encoding classes proposed by \cite{agile:2019} are extended to capture events and event types, as well as dates and numbers. Logical operations such as \textit{and}, \textit{or}, \textit{negation} and \textit{comparison} are then added to the encoding schema. Table~\ref{tab:encodings} shows the schema of encoding classes and their codes. Identifying the extended encodings enable us to support the identified concepts in the object-based conceptualization of place (i.e., location, place and event), and also support complex questions that contain multiple criteria.

\begin{table}[]
  \caption{Encoding classes}
  \label{tab:encodings}
  \centering
    \begin{tabular}{cl|cl}
    \toprule
        Encoding class & Code & Encoding class & Code\\
    \midrule
        \ex{where} & \pattern{1} & \ex{place name} & \pattern{P}\\
        \ex{what} & \pattern{2} & \ex{place type} & \pattern{p} \\
        \ex{which} & \pattern{3} & \ex{event} & \pattern{E} \\
        \ex{when} & \pattern{4} & \ex{event type} & \pattern{e}\\
        \ex{how} & \pattern{5} & \ex{properties} & \pattern{o}\\
        \ex{how+adj} & \pattern{6} & \ex{activity} & \pattern{a}\\
        \ex{why} & \pattern{7} & \ex{situation} & \pattern{s}\\
        \ex{is/are} & \pattern{8} & \ex{spatial relation} & \pattern{R}\\
        \ex{date} & \pattern{d} & \ex{temporal relation} & \pattern{r}\\
        \ex{place quality} & \pattern{Q} & \ex{properties/events quality} & \pattern{q}\\
        \ex{comparison} & \pattern{<}, \pattern{>}, \pattern{=} & \ex{and} & \pattern{\&}\\
        \ex{or} & \pattern{|} & \ex{negation} & \pattern{!}\\
    \bottomrule
    \end{tabular}
\end{table}

To extract the encodings, the pre-trained models for fine-grained named entity recognition (NER) \cite{Lample2016NeuralAF} and part-of-speech tagging \cite{Joshi2018ExtendingAP} are used. The relation between the encoding classes and part-of-speech is then used to extract information from the questions as follows:
\begin{itemize}
    \item \textbf{Noun encoding:} Noun phrases can be place names, place types, event names, event types or properties. Place names and event names are detected using a fine-grained NER \cite{lample2016neural}. Generic nouns are identified using part-of-speech tagging. These noun phrases include place types and event types. Here, a look-up approach is used to test whether a generic noun refers to a place type or an event type. The lists of place types and event types are extracted from the OpenStreetMap tags and the YAGO ontology, respectively. The remaining unlabelled noun phrases are labelled as properties. 
    
    \item \textbf{Verb encoding:} As proposed and tested by \cite{agile:2019}, sets of active and stative verbs can be used to differentiate situations from activities. Here, BERT representations of the words are used to derive cosine similarity of any identified verb to predefined sets of active and stative verbs \cite{agile:2019}, and classify them as situations or activities based on maximum similarity.
    
    \item \textbf{Preposition encoding:} Preposition phrases can be related to spatial relations, temporal relations, and comparisons. If a preposition refers to place(s) (either by place names or \emph{generic} place types), the preposition is captured as a candidate for spatial relations. If the preposition phrase refers to a date then the preposition is captured as a temporal relation. Otherwise, if the preposition phrase includes \textit{comparative adjectives}, it is captured as a comparison (e.g., \emph{greater than}). Here, constituency parsing \cite{Joshi2018ExtendingAP} is used to find what phrase a preposition refers to.
    
    \item \textbf{Adjective encoding:} 
    Superlative (e.g., \emph{smallest}) and descriptive (e.g., \emph{small}) adjective phrases that are referring to place names and place types are captured as place qualities, and otherwise they are captured as property qualities. Comparative adjectives are a constituent of preposition phrases (e.g., \emph{smaller than}) which are encoded as comparisons due to their different role in the questions.
    
    \item \textbf{Conjunction encoding:} Conjunctions are identified through part-of-speech tagging, and their encodings are identified through a look-up approach. The conjunctions are encoded either as \textit{or}, \textit{and}, or \textit{negation}.
\end{itemize}

\subsection{Constituency and Dependency Parsing}
\label{sec:intent}
\begin{figure}[t]
\includegraphics[width=1\linewidth] {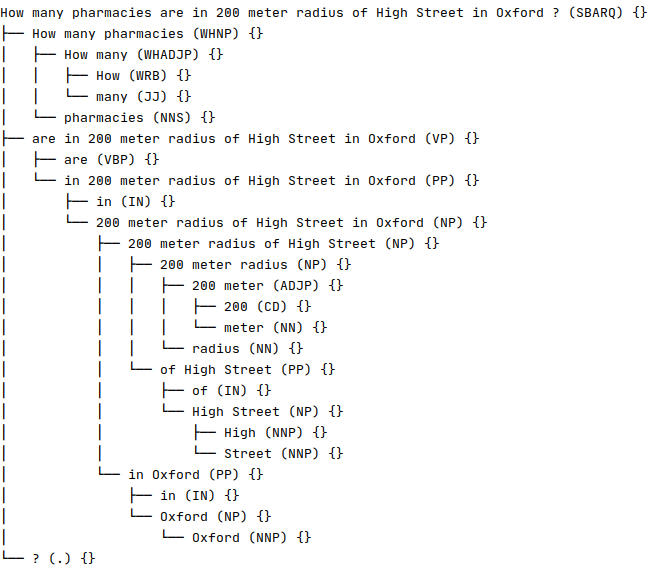}
\caption{Constituency parsing}
\label{fig:c_parsing}
\end{figure}

\begin{figure}[t]
\centering
\includegraphics[width=0.65\linewidth] {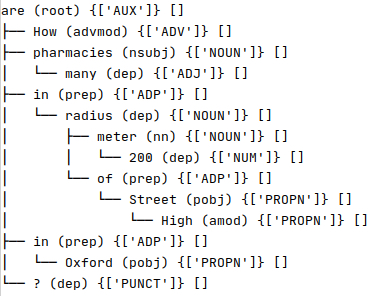}
\caption{Dependency parsing}
\label{fig:d_parsing}
\end{figure}

Constituency parsing and dependency parsing are two approaches to study the structure of natural language sentences. Constituency parsing \cite{Joshi2018ExtendingAP} captures how tokens can be combined to construct phrases and how phrases form more complex phrases or sentences (Figure~\ref{fig:c_parsing}). Dependency parsing \cite{DozatM17} identifies relations between tokens and captures their long distance relations (Figure~\ref{fig:d_parsing}). In our method, if a relation between identified concepts (e.g., places and events) can be extracted from phrases and their constituents, constituency parsing is used; otherwise we use dependency parsing to make sure the long distance relations are captured.

\begin{figure}[t]
\includegraphics[width=1\linewidth] {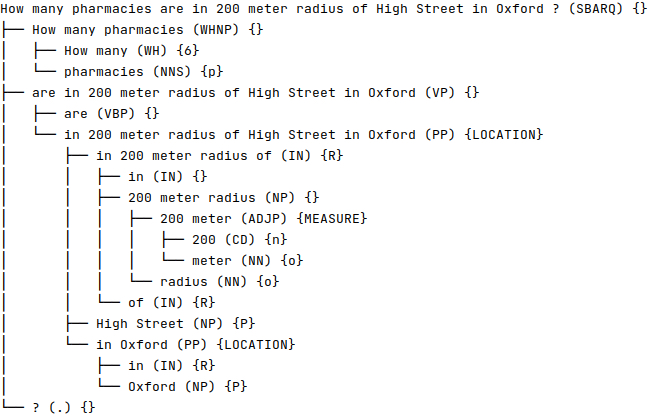}
\caption{Labelled constituency parse tree}
\label{fig:c_parsing_labelled}
\end{figure}

The parsing workflow starts with constituency parsing and identification of phrase-level information such as \textit{conjunctions phrases}, \textit{quality phrases} and \textit{location phrases}. The following steps are performed in analysing constituency parsing results:

\begin{enumerate}
    \item \textbf{Preprocessing:} In this step, the constituency parse tree is trimmed and encoded. During trimming, extracted compound phrases (e.g., High Street) are captured as leaf nodes in the tree representation. Hence, the constituency tree expands to a meaningful phrase level that is not necessarily consisting of individual tokens. Then, the extracted encodings are labelled in the tree representation.
    \item \textbf{Conjunction phrases:} First, leaf nodes labelled with conjunction encodings (i.e., \pattern{\&}, \pattern{|}, and \pattern{!}) are selected from the constituency tree. For \textit{and}/\textit{or} conjunctions, if the parent node includes constituents of the same encoding class (e.g., multiple place names), the parent node is labelled as a \textit{conjunction phrase} -- e.g., towns or cities in \textit{what are the towns or cities in UK?}. For negations, if the parent nodes contain place names or event names the parent nodes are labelled as \textit{negation phrase} -- e.g., `except London' in \textit{What is the largest city in UK except London?}
    \item \textbf{Quality phrases:} Leaf nodes encoded as qualities (i.e., \pattern{q} and \pattern{Q}) are retrieved from the constituency tree. If their parent nodes include nodes with relevant encoding classes (i.e., \pattern{p}, \pattern{P}, \pattern{e}, \pattern{E} and \pattern{o}), the parent nodes are captured as quality phrases.
    \item \textbf{Location phrases:} Location phrases are constructed by spatiotemporal relations and their corresponding anchor places/dates -- e.g., \emph{in 200 meter radius of High Street}. Here, location is considered in space and time. Location phrases are captured by finding an ancestor node of spatiotemporal relations which includes the anchor places and dates. Figure~\ref{fig:c_parsing_labelled} shows two labelled locations, \textit{in Oxford} and \textit{in 200 meter radius of High Street in Oxford}.
    \item \textbf{Measure Phrases:} Phrases that are only constructed with cardinal numbers and properties/types are detected as measures phrases. In the introductory example, \textit{200 meter} is a phrase with numeric (200) and property (meter) constituents. 
    
    \item \textbf{Comparison phrases:} Comparison is a binary relation with a source and a target. While the target and comparison tokens (e.g., \textit{less than}) often form a phrase, the source can have a long distance relation with comparison tokens, depending on the structure of the questions. Hence, the relation between comparison tokens and the target is captured through constituency parsing, and later dependency parsing is used to detect the source of comparison. If the parent node of an identified comparison encoding includes a valid target phrase (i.e., places, events, properties, dates or measure phrases), the parent node is labelled as a comparison phrase (e.g., \emph{more than ten districts}).
\end{enumerate}
The results of detecting phrase-level information is shown in the labelled constituency tree in Figure~\ref{fig:c_parsing_labelled}.

Next, we extract the \textit{intent} of the place-related questions using the following heuristics:
\begin{enumerate}
    \item \textbf{Question word rule:} The question word gives an initial signal about the intent of the questions -- e.g., \ex{where}, \ex{what} and \ex{is/are}. For yes/no questions (\ex{is/are}) not only intent is clear but the answer domain -- a Boolean value -- is evident as well. For other question words, we need to identify the concept that is the intent of the question. For \ex{where} questions, the intent is a location of a place or an event. \textit{How+adjective} question words determine specific operations such as counting (e.g., how many) and distance (e.g., far) that must be applied to the intent concepts.
    \item \textbf{Specificity rule:} The more specific the concepts are, the less likely they are to be the intent of the question --- e.g., a \textit{property} such as population or \textit{place/event type} (e.g., \textit{cafe}) is more likely to be the intent of a question in comparison to a \textit{place/event name}. For example, in the question \ex{Which cities in England have at least two castles?}, \textit{cities} and \textit{castles} are more likely to be the intent of the question than \textit{England}. This rule determines the specificity of concepts using their encoding classes, i.e., \textit{properties}, \textit{place/event types} and \textit{place/event names}. The properties are the least specific concept. The next more specific concept is \textit{place/event type}, and finally the most specific concept is \textit{place/event name}. 
    \item \textbf{Phrase rule:} This rule determines whether a candidate concept is valid to be considered as the intent concept. If the candidates belong to place types or place names (e.g., multiple place types are identified), location phrases are used to reduce ambiguity. If a place type or place name is the intent, it must not belong to a \textit{location phrase}. For example, in \textit{Where in the UK is Wolverhampton?}, the UK is part of a location phrase (\textit{in the UK}) and must be removed from the list of intent candidates. Such location phrases are spatial criteria, and must be avoided in the intent recognition process. In the same manner, properties which belong to activity/situation phrases, complex spatial relations and comparison phrases are removed from the candidate list.
    \item \textbf{Phrase position rule:} Finally, if the intent is still ambiguous, and multiple valid concepts are identified, a  position rule is used, favouring concepts appearing earlier in the question phrase. The subject of the questions is then the earliest concept extracted from the questions. For example, in the question: \ex{Which river crosses the most cities in England?}, the earliest concept (\textit{river}) is determined as the intent.
\end{enumerate}
In the introductory example, the intent is therefore identified as \textit{How many pharmacies} using these rules.

\begin{figure}[t]
\includegraphics[width=0.65\linewidth] {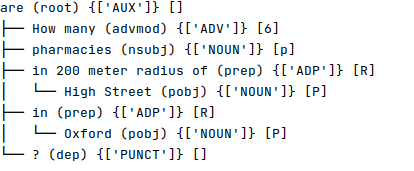}
\caption{Dependency parse tree after preprocessing}
\label{fig:d_parsing_labelled}
\end{figure}

Next, dependency parsing is used to detect the relation between (1) places/events with location phrases, (2) situation/activities with properties, (3) places with situations/activities, and (4) comparison phrases and their source.
\begin{enumerate}
    \item \textbf{Preprocessing:} The dependency parse tree is trimmed and enriched by the extracted encoding and the identified phrases from constituency parsing. In the trimming task, each of the extracted compound phrases (e.g., \emph{High Street}) or phrase-level information from constituency parsing is captured as a single node in the dependency tree representation. Figure~\ref{fig:d_parsing_labelled} shows the dependency tree of the introductory example after the preprocessing step.
    \item \textbf{Places/events and location phrases:} Using the terminology introduced by \cite{RN38}, a spatial description includes three elements, the \textit{locatum}, the \textit{spatial relation}, and the \textit{relatum}. A location phrase is a combination of the spatial relation and the relatum that describe the location of a place or an event in space and time (e.g., \textit{in 200 meter radius of High Street}). If the location phrase is a dependent (\textit{prep} relation\footnote{more information can be found at: \url{https://universaldependencies.org/u/dep/}}) of a place or an event, their relation is captured as a locatum and location phrase relation (e.g., \textit{pharmacies} (locatum), \textit{in 200 meter radius of High Street} (location phrase)).
    \item \textbf{Situations/activities with properties:} Situations and activities often include references to non-place objects that are captured as properties. In these cases, the verbs do not completely describe the situation and activities -- e.g., to buy [coffee]. If a property is a dependent (an object phrase of the verb, \textit{dobj} relation) of a situation/activity verb, then the relation is identified between the verb and property phrase. 
    \item \textbf{Places with activities/situations:}
    If a situation/activity phrase is a dependant (subject phrase of the verb, \textit{nsubject} relation) of an identified place, their dependency is captured as a place relation with situations/activities. The same grammatical rule is applied for events and situation verbs (e.g., [...] \textit{hurricanes occurred} [...]).
    \item \textbf{Source of comparison phrases:} If a comparison phrase is a direct dependent of a valid encoding class (i.e., \pattern{p}, \pattern{P}, \pattern{e}, \pattern{E} and \pattern{o}), their relation is captured. In \textit{does England have more counties than Ireland}, England has a valid encoding class (i.e., \pattern{P}) and the comparison phrase (i.e., \textit{more counties than Ireland}) is its dependent in the parse tree.
\end{enumerate}

\subsection{Generating Logical Statements}
\label{section:fol}
Logical statements are constructed using \textit{terms} and \textit{functions}. In place-related QA, the terms are places, events or their properties. A term can be either a constant (e.g., \textit{High Street}) or a variable (e.g., \textit{x} that represents pharmacies). Place and event names are the constants and the types and properties are the variables terms. Referring to the introductory example, \textit{High Street} and \textit{Oxford} are the constants, and for the \textit{pharmacies}, variable $x0$ is assigned to represent the generic reference. Functions are symbols that either declare terms or describe their relations -- e.g., declaration: ${Place}(High Street)$ and relation: $IN(High Street, Oxford)$. The logical statements are either query statements that return values of variable terms or Boolean True/False statements.

We declare terms using two special functions, $PLACE()$ and $EVENT()$, that describe place and event concepts. Generic references are declared using the assigned variables and the extracted type -- e.g., $PHARMACY(x0)$. A generic declaration is either a place or an event using a predefined rule, e.g., $\forall x\: PHARMACY(x) \rightarrow PLACE(x)$.

Spatiotemporal relations, situation/activity relations, comparisons, and qualities are defined as functions. Qualities describe a single term, and are represented by a function with an argument. 
Spatial relations are either binary or ternary functions. The name of the functions are the spatial prepositions, with the identified \textit{locatum} and \textit{relatum} as the two arguments. In case of complex spatial relations, the additional information is presented as the third argument. In the example, the complex relation is represented as $IN\_RADIUS\_OF(x0,\ High\ Street,\ 200\ meter)$.
Comparisons are also defined as binary relations with the comparison source as the first argument and target as the second argument.

The extracted conjunction relations are treated differently when generating logical statements. If a constituent of a conjunction participates in relations to other terms, then automatically a similar relation is applied to the other constituent of the conjunction. For example in \textit{Are there any rivers that cross both England and Wales?}, the spatial relations between the rivers and \textit{England} is replicated for \textit{Wales} because of their conjunction relations -- i.e., $CROSS(x0, England) \:\land\: CROSS(x0, Wales)$.

The logical representation is formulated using the identified intent and concatenating declarations and functions. If the question is a yes/no question, then the statement is simply derived by appending term declarations and function definitions. Otherwise, the logical statement is a query statement that returns the intended terms (e.g., $x0$) or functions (e.g., $COUNT(x0)$). The logical statement for the introductory question is presented as:
\begin{equation}
\begin{longformulas}
COUNT(x0): PLACE(High\ Street) \:\land\: PLACE(Oxford) \:\\
\land\: PHARMACY(x0) \:\land\: IN\_RADIUS\_OF(x0, High\ Street, 200\:meter) \\ \:\land\: IN(High\ Street, Oxford)     
\end{longformulas}
\end{equation}

\subsection{Generating GeoSPARQL Queries}
To generate GeoSPARQL queries, two necessary steps are needed: 1) concept identification and ontology mapping, to match extracted information to available information in the knowledge base; and 2) the dynamic generation of GeoSPARQL queries.

\textbf{Concept Identification and Ontology Mapping}:
Concept identification is completed prior to query generation. We use Apache Solr to index the names and identifiers of places and events in the knowledge base and perform string matching using Solr search. Solr improves the performance of concept identification through its powerful string indexing. While concept identification could be undertaken on the query itself, string matching can take long on large knowledge bases.

We consider a one-to-many mapping to match extracted place/event types and properties to the knowledge base ontology. Our workflow performs ontology mapping in three sequential steps: 1) an \textit{exact matching} from extracted information to the knowledge base ontology, followed by 2) a \textit{label matching} using cosine similarity between the contextual BERT representations \cite{kenton2019bert} of the extracted information from the questions and the labels in our ontology; and finally 3) a \textit{glossary matching} using cosine similarity between the BERT representations of the definitions for the extracted information from WordNet and Wikipedia snippet search with glossary in our ontology. Both label and glossary matching are based on thresholds that are tuned using randomly selected and manually mapped types and properties (10\% of the available set).

\textbf{Query Generation}:
GeoSPARQL queries have several constituents, including \textit{PREFIXES}, \textit{ASK/SELECT} statements, and \textit{WHERE} clauses. PREFIXES define the namespaces from which to access knowledge bases, ontologies, and implemented functions. We use a set of predefined prefixes that includes GeoSPARQL functions, YAGO and YAGO2geo ontologies and resources. The ASK/SELECT statements determine the output of the query, and the WHERE clauses capture the criteria mentioned in the question.

We propose three sequential steps to translate logical statements to GeoSPARQL queries: 1) the overall structure of a query (i.e., ASK vs.\ SELECT query) is determined from the extracted intent. In case of SELECT queries, the intent of questions determines what variables are queried for their values; 2) the WHERE-clause is dynamically generated by concatenating individual concept and relation definition statements; and finally 3) sorting and aggregation (ORDER-BY and GROUP-BY clauses) are generated for queries that require these. 
Details of the predefined templates are presented in Appendix \ref{app:templates}.

The WHERE-clause is part of the general structure for both ASK and SELECT queries. To generate the WHERE-clause, a unique variable is assigned to each extracted geographic concept. Each of these concepts is defined using the predefined templates and a corresponding variable name. The concept definition statements define a place/event based on the \textit{name} or \textit{type}. The extracted \textit{properties} are defined based on their corresponding place/event variables. The extracted relations among the concepts (e.g., \textit{spatiotemporal relations}) are translated to GeoSPARQL query using the participating variables, and the predefined relation templates.

Finally, when aggregation (e.g., counting) or sorting (e.g., superlative qualities) is needed their corresponding templates are used. In the special case of aggregation and sorting, GROUP-BY and HAVING statements, and ORDER-BY and LIMIT-statements are concatenated at the end of the generated query, respectively. Sorting is needed when superlative qualities (e.g., longest river) are found, and aggregation is necessary when comparison-based situations (e.g., cities that have more than 10 suburbs) are extracted. The generated GeoSPARQL query for the introductory example is shown in Query~\ref{lst:exsparql} (see Appendix \ref{app:templates}).

\section{Results and Discussion}
\label{sec:chap_5_res}

\subsection{Extraction Results and Logical Statements} 
Table~\ref{tab:results_encodings} shows the results of evaluating encoding extraction using Geospatial Gold Standard dataset \cite{Punjani:2018}. The macro averaging strategy is used to derive the precision, recall and f-score of encoding extraction for each encoding class. The count shows the frequency of each class in the question dataset based on manual annotation.

As shown in Table~\ref{tab:results_encodings}, the pre-trained models perform well in place name identification, and the part-of-speech rules are successful in identifying numbers, comparisons and conjunctions. The dependency parsing and constituency parsing are highly successful in identifying more complex encoding classes such as spatiotemporal relations and qualities. 

Table~\ref{tab:results_encodings} shows that event-based place-related questions are completely missing in the dataset. Moreover, activities are also rarely observed in the dataset, which is another limitation of the Geospatial Gold Standard dataset \cite{Punjani:2018}. Identifying such limitations shows that by using the object-based conceptualization our method is less prone to the biases in the dataset.

\begin{table}[]
  \caption{Extraction performance}
  \label{tab:results_encodings}
  \centering
  \resizebox{\linewidth}{!}{
    \begin{tabular}{lcccc}
    \toprule
        Encoding class & Average precision & Average recall & Average f-score & Count\\
    \midrule
    \ex{Place name} & 98.9 & 99.5 & 99.2 & 263\\
    \ex{Place type} & 100.0 & 99.4 & 99.7 & 178\\
    \ex{Event name} & -- & -- & -- & 0\\
    \ex{Event type} & -- & -- & -- & 0\\
    \ex{Properties} & 95.9 & 97.9 & 96.9 & 55\\
    \ex{Number}& 100.0 & 100.0 & 100.0 & 26\\
    \ex{Spatiotemporal relation} & 100.0 & 96.8 & 98.4 & 191\\
    \ex{Situation} & 94.2 & 98.0 & 96.1 & 51\\
    \ex{Activity} & 50.0 & 100.0 & 66.7 & 1\\
    \ex{Place quality} & 100.0 & 91.7 & 95.7 & 25\\
    \ex{Object quality} & 85.7 & 100.0 & 92.3 & 6\\
    \ex{Comparison} & 100.0 & 100.0 & 100.0 & 24\\
    \ex{Conjunction} & 100.0 & 100.0 & 100.0 & 6\\
    \bottomrule
    \end{tabular}
    }
\end{table}

Table~\ref{tab:results_fol} shows the average precision, recall and f-score of logical term declarations and function definitions. The results show that declarations, spatiotemporal relations and comparisons can be formulated in logical statements with high precision and recall. Thus, the combination of grammatical rules from dependency parsing and constituency parsing are shown to be successful.

Table~\ref{tab:results_fol} shows that qualities are defined with high precision (100\%), yet the recall is lower (82.8\%). The reason is that using constituency parsing alone may not be sufficient for all cases. For example in \textit{Which site of Manchester is the most popular?}, the long distance relation between the quality (\textit{most popular}) and the place type it describes (\textit{site}) cannot be captured through constituency parsing. Similarly for situations, we observe higher precision and lower recall. The reason is that in some cases dependency parsing fails to link situation verbs to properties. Hence, when the link is not captured, the situation is missing and the recall is lower.

Conjunctions are reflected in the logical statements with high recall (i.e., 100\%) and lower precision (i.e., 80\%). The conjunctions are reflected in logical statements by applying the same functions for both sides of the conjunction which are bounded by either \textit{and} or \textit{or} logical operators. Hence, errors in defining other functions (e.g., spatial relations) can be propagated through conjunctions. Consequently, errors in these functions impact on the precision of the conjunction statements as well.

\begin{table}[t]
  \caption{Evaluation of the generated logical statements}
  \label{tab:results_fol}
  \centering
  \resizebox{\linewidth}{!}{
    \begin{tabular}{lccc}
    \toprule
        Term/Function/Statement & Average precision & Average recall & Average f-score\\
    \midrule
    \ex{Declaration} & 98.2 & 99.7 & 98.9 \\
    \ex{Spatiotemporal relations} & 97.6 & 91.9 & 94.7 \\
    \ex{Quality} & 100.0 & 82.8 & 90.6\\
    \ex{Comparison} & 91.7 & 91.7 & 91.7 \\
    \ex{Conjunction} & 80.0 & 100.0 & 88.9\\
    \ex{Situation} & 100.0 & 64.7 & 78.6 \\
    \ex{Overall}\footnote{Each logical statement is evaluated either as a correct or incorrect representation. Hence, the recall and fscore are meaningless for such evaluation} & 85.0 & -- & --\\
    \bottomrule
    \end{tabular}
    }
\end{table}

\subsection{Query Generation Results}
Table~\ref{tab:query_results} shows the results of analyzing concept identification, ontology mapping and query generation. The concept identification is highly accurate, yet in this evaluation we did not consider toponym disambiguation. Hence, the minor issues in concept identification is due to incorrect or missing data in our knowledge base (e.g., an incorrect alternative name for a place). For questions that contain spatial relations between two place names, toponym disambiguation is automatically resolved when the query is executed. However, if we only have one place name in the question, every place with the same name may be considered as a correct match -- simply, because of lack of additional context.

\begin{table}[t]
  \caption{Query generation results}
  \label{tab:query_results}
  \centering
  \resizebox{\linewidth}{!}{
    \begin{tabular}{lccc}
    \toprule
    Step&Average precision&Average recall&Average f-score\\
    \midrule
    \ex{Concept identification}&97.5&96.3&96.9\\
    \ex{Ontology mapping}&83.7&97.2&89.9\\
    \ex{Query} & 79.5 & -- & --\\
    \bottomrule
    \end{tabular}
    }
\end{table}

The results show that the ontology mapping method using BERT embedding performs well in matching extracted information to the predefined ontology. However, the evaluation of ontology mapping is sometimes subjective, specially when multiple knowledge sources are integrated. For example, whether `clinic' can be a valid match for `hospital' is subjective. Here, if the services associated to the place types are similar, the match is considered correct. Yet, further larger-scale evaluation studies are needed to measure human agreement on this task.

We identify the following reasons for the precision drop when generating queries from logical statements (drop from 85\% to 79.5\%)\footnote{Check Appendix \ref{app:geosparql} for the evaluation of generating GeoSPARQL constituents.}:
\begin{itemize}
    \item Flexibility of natural language: For example, \textit{Scottish counties} can be captured in logical form, yet the adjective implies a spatial relation which must be stated in the corresponding GeoSPARQL query -- i.e., \textit{counties of Scotland}. In such cases, our rule-based approach fails to properly generate the query.
    \item Missing concepts and types: In some cases the place names and types that are mentioned in the questions are not found in the knowledge base, and thus query generation fails -- e.g., \textit{underground lines} cannot be matched to a place type in our knowledge base.
\end{itemize}


\subsection{Comparing with Previous Works}
Table~\ref{tab:comp_res} shows the results of analysing generated GeoSPARQL queries in comparison to previous work. To the best our knowledge, three papers have proposed methods for translating questions to GeoSPARQL and used the same dataset. Table~\ref{tab:comp_res} shows a considerable improvement on the benchmark dataset in comparison to previous works \cite{Punjani:2018, punjani2021templatebased,li2021ur}.

In terms of the number of questions that can be handled, the methods presented by \cite{Punjani:2018} and \cite{punjani2021templatebased} cover 43\% and 76.5\% of the questions, respectively. Using dynamic query generation, our approach and \cite{li2021ur} were able to generate queries for all of the questions in the dataset. The precision of the answers is only reported by \cite{punjani2021templatebased, Punjani:2018}, and the comparison shows an improvement of 14\% of precision in retrieving the answers, attributable to our method.

\begin{table}[t]
	\centering
	\caption{Query generation results. ``*'' indicates that the results are only for correct queries, not the whole dataset.}
	 \resizebox{\linewidth}{!}{
	\begin{tabular}{ccccc}
		\toprule
		& Template coverage & Correct query & Answers\\
		\midrule
		GeoQA \cite{Punjani:2018} & 43.0 & 22.0 (51.2\% of 43\% generated query) & 37.4*  \\
		NeuralGQA \cite{li2021ur} & 100.0 & 38.0 (71\% of 45\% generated query) & -- \\
		GeoQA \cite{punjani2021templatebased} & 76.5 & 65.5\% & 64.6*  \\
		Our approach & \textbf{100.0} & \textbf{79.5} & \textbf{78.6*} \\
		\bottomrule
	\end{tabular}
	\label{tab:comp_res}
	}
\end{table}

The comparison shows that integrating dependency parsing and constituency parsing is useful to identify the relations. The benchmark methods \cite{Punjani:2018,punjani2021templatebased} only use dependency parsing to extract information from the questions. Hence, using phrase level information from constituency parsing can improve the quality of information extraction.

Using the object-based conceptualization \cite{purves:2019}, our method covers more diverse types of place-related questions and consequently is able to handle all questions in the dataset. 
Moreover, the proposed logical representation has two main advantages which ease such translations. First, the representation is machine digestible and can also be empowered with logical reasoning. Second, it can capture both intent and criteria of the natural language questions which is also required in translating natural language questions to other structured query language(s).

\section{Conclusions}
In this paper, we present a method to translate place-related questions to queries using the object-based conceptualization of place \cite{purves:2019}. Using the domain knowledge, the proposed method is less biased to the dataset and covers more diverse types of place-related questions, and we identified missing types of questions in the Geospatial Gold Standard dataset \cite{Punjani:2018} (i.e., questions about events and activities).

In our method, we use and test state-of-the-art pre-trained models, and the results shows that the quality of information extraction and relation identification is improved in comparison to the benchmark methods. Moreover, the grammatical rules derived from the dependency and constituency parsing lead to more accurate results in digesting place-related questions. However, the available dataset is relatively small and the method should be tested whenever larger datasets are available for GeoQA. Enriching the current benchmark dataset to cover questions about activities and events is also a necessary step for future work in translating place-related questions to queries. Using local context (e.g., user location) to present relevant and personalized answers to geographic questions remains as a future work of this study.

\section*{Acknowledgement}
The support by the Australian Research Council grant DP210101156 is acknowledged.

\nocite{*}


\appendix
\section{Predefined templates}
\label{app:templates}
Definition statements for places using their \textit{name} and \textit{type} are presented in Queries~\ref{lst:name} and \ref{lst:type}, respectively. Here, \textit{PI} refers to variable name for place identifiers, and \textit{URIS} refers to the results of concept identification in Query~\ref{lst:name} and ontology mapping in Query~\ref{lst:type}. In place declarations, the geometry is always expressed using the Well-Known Text (WKT) encoding (\textit{?<PI>GEOM}).

\begin{minipage}{1\linewidth}
\begin{lstlisting}[captionpos=b, caption={Place definition template using name}, label={lst:name},
   basicstyle=\scriptsize \ttfamily,frame=single]
VALUES ?<PI>  {<URIS>}.
?<PI> geosparql:hasGeometry ?<PI>G .
<PI>G geosparql:asWKT ?<PI>GEOM .
\end{lstlisting}
\end{minipage}

\begin{minipage}{1\linewidth}
\begin{lstlisting}[captionpos=b, caption={Place definition template using type}, label={lst:type},
   basicstyle=\scriptsize \ttfamily,frame=single]
?<PI> rdf:type ?<PI>TYPE;
    geosparql:hasGeometry ?<PI>G .
?<PI>G geosparql:asWKT ?<PI>GEOM .
VALUES ?<PI>TYPE {<URIS>} . 
\end{lstlisting}
\end{minipage}

Properties are captured through \textit{has-a} relations by binding identified properties to the results of ontology mapping (see Query~\ref{lst:attribute}). In the case of spatial relations, a lookup table is used to map spatial prepositions to their corresponding spatial functions implemented in GeoSPARQL. The relations are constructed by the \textit{predicate} and \textit{argument(s)} which are captured in the corresponding function definition. Similarly, the situation/activities and comparisons are defined using their corresponding templates from their definitions. The template for distance relations, as examples of the relation templates, is shown in Query~\ref{lst:distance}.

\begin{minipage}{1\linewidth}
\begin{lstlisting}[captionpos=b, caption={Attribute relation template}, label={lst:attribute},
   basicstyle=\scriptsize \ttfamily,frame=single]
VALUES ?<ATTRIBUTE> {<URIS>}.
?<PI> ?<ATTRIBUTE> ?<PROPERTY>.
\end{lstlisting}
\end{minipage}

\begin{minipage}{1\linewidth}
\begin{lstlisting}[captionpos=b, caption={Distance relation template}, label={lst:distance},
   basicstyle=\scriptsize \ttfamily,frame=single]
FILTER(geof:distance(?<PI1>GEOM, ?<PI2>GEOM, <UNIT>) < <DISTANCE>).
\end{lstlisting}
\end{minipage}

The template for aggregation and sorting are shown in Queries~\ref{lst:aggregation} and \ref{lst:sorting}, respectively.

\begin{minipage}{1\linewidth}
\begin{lstlisting}[captionpos=b, caption={Aggregation template}, label={lst:aggregation},
   basicstyle=\scriptsize \ttfamily,frame=single]
GROUP BY <VARIABLE>
HAVING (COUNT(*) > <LIMIT>)
\end{lstlisting}
\end{minipage}

\begin{minipage}{1\linewidth}
\begin{lstlisting}[captionpos=b, caption={Sorting template}, label={lst:sorting},
   basicstyle=\scriptsize \ttfamily,frame=single]
ORDER BY <ASC/DESC> (<VARIABLES>) LIMIT <LIMIT>.
\end{lstlisting}
\end{minipage}

\begin{minipage}{1\linewidth}
\begin{lstlisting}[captionpos=b, caption={GeoSPARQL query of the introductory example}, label={lst:exsparql},
  basicstyle=\scriptsize \ttfamily,frame=single]
PREFIX geosparql: <http://www.opengis.net/ont/geosparql#>
PREFIX geof: <http://www.opengis.net/def/function/geosparql/>
PREFIX units: <http://www.opengis.net/def/uom/OGC/1.0/>
PREFIX rdf: <http://www.w3.org/1999/02/22-rdf-syntax-ns#>
PREFIX yago:<http://yago-knowledge.org/resource/>
PREFIX geont: <http://kr.di.uoa.gr/yago2geo/ontology/>
PREFIX yago2geo:<http://kr.di.uoa.gr/yago2geo/resource/>
SELECT DISTINCT (COUNT(distinct ?x0) as ?countx0) 
WHERE { 
VALUES ?c0 {yago2geo:OSM_HighStreet246
            yago2geo:OSM_HighStreet301}. 
?c0 geosparql:hasGeometry ?c0G .
?c0G geosparql:asWKT ?c0GEOM .
VALUES ?c1  {yago:Oxford
            yago2geo:OSM_Oxford813}. 
?c1 geosparql:hasGeometry ?c1G .
?c1G geosparql:asWKT ?c1GEOM .
?x0 rdf:type ?x0TYPE;
    geosparql:hasGeometry ?x0G .
?x0G geosparql:asWKT ?x0GEOM .
VALUES ?x0TYPE {yago:wordnet_drugstore_103249342
                geont:OSM_amenity_veterinary
                geont:OSM_amenity_pharmacy} . 
FILTER(geof:distance(?x0GEOM, ?c0GEOM, units:meter) < 200).
FILTER (geof:sfContains(?c1GEOM, ?x0GEOM)). }
\end{lstlisting}
\end{minipage}

\section{Evaluation of GeoSPARQL components}
\label{app:geosparql}
The performance for generating each part of the GeoSPARQL queries (e.g., WHERE clause) is presented in Table~\ref{tab:results_geosparql}. All GeoSPARQL queries include intent (SELECT/ASK statement) and criteria (WHERE clause). Hence, the recall and f-score is meaningless in evaluating these parts (i.e., 100\% recall). On the other hand, sorting (29 questions) and aggregation (6 questions) are applicable for specific questions, where recall and f-score are reported (Table~\ref{tab:results_geosparql}). 

Table~\ref{tab:results_geosparql} shows that the intent heuristic performs well and therefore the errors in formulating the criteria (WHERE clause) are the main source of incorrect GeoSPARQL queries. Criteria may include multiple components, and even a minor mistake in formulating a criterion in a WHERE clause produces an incorrect query.

\begin{table}[htb]
  \caption{GeoSPARQL queries evaluation}
  \label{tab:results_geosparql}
  \centering
  \resizebox{\linewidth}{!}{
    \begin{tabular}{lccc}
    \toprule
        GeoSPARQL query & Average precision & Average recall & Average f-score\\
    \midrule
    \ex{Intent (SELECT/ASK)}  & 97.5 & -- & --\\
    \ex{Criteria (WHERE)} &	84.3 & -- & -- \\
    \ex{Aggregation (Group By)}& 100.0 & 83.3 & 91.5\\
    \ex{Sorting (Order By)} &	92.9 & 89.7 & 91.2\\
    \bottomrule
    \end{tabular}
    }
\end{table}
\end{document}